\documentclass[a4paper,12pt]{article}
\usepackage{amsmath}
\usepackage{accents}
\usepackage[english]{babel}
\usepackage{graphicx}
\usepackage[usenames]{color}
\usepackage{colortbl}

\input epsf.tex

\begin{document}

\centerline {\bf The quantum brachistochrone problem for an arbitrary spin in a magnetic field}
\medskip
\centerline {A. R. Kuzmak$^1$, V. M. Tkachuk$^2$}
\centerline {\small \it E-Mail: $^1$andrijkuzmak@gmail.com, $^2$voltkachuk@gmail.com}
\medskip
\centerline {\small \it Department for Theoretical Physics, Ivan Franko National University of Lviv,}
\medskip
\centerline {\small \it 12 Drahomanov St., Lviv, UA-79005, Ukraine}

{\small

We consider quantum brachistochrone evolution for a spin-$s$
system on rotational manifolds. Such manifolds are determined by
the rotation of the eigenstates of the operator of projection of
spin-$s$ on some direction. The Fubini-Study metrics of these
manifolds are those of spheres with radii dependent on the value
of the spin and on the value of the spin projection. The
conditions for optimal evolution of the spin-$s$ system on
rotational manifolds are obtained.

\medskip

}

\section{Introduction\label{sec1}}

The interest in quantum brachistochrone problem has increased
after the publication of paper \cite{QBP}, where Carlini et al.
considered the following problem: What is the optimal Hamiltonian,
under a given set of constraints, such that the evolution from a
given initial state $\vert\psi_i\rangle$ to a given final one
$\vert\psi_f\rangle$ is achieved in the shortest time? Using a
variational principle, the authors of this work solved the
brachistochrone problem for some specific examples of constraints.
In \cite{OHfST}, results analogous to those of \cite{QBP} were
obtained more directly due to symmetry properties of the quantum
state space. The authors of this paper showed that the
brachistochrone evolution between two states $\vert\psi_i\rangle$
and $\vert\psi_f\rangle$ which are set in the Hilbert space of
dimension $n$ is reduced to the evolution on the two-dimensional
subspace spanned by the two vectors $\vert\psi_i\rangle$ and
$\vert\psi_f\rangle$.

It is easy to find a geodesic between two quantum states and
optimal Hamiltonian for a two-level system with a given set of
constraints. The quantum brachistochrone problem for a such system
was considered in many papers (see, for example, \cite{ref1, ref2,
ref3, FSM}). There are multilevel physical systems with
dimensionality higher than two whose properties do not allow
reducing of quantum evolution to the evolution on the
two-dimensional subspace as in the paper \cite{OHfST}. For
example, the Hamiltonian of a spin-$s$ system (where $s>1/2$) in
an external magnetic field contains only two free parameters,
which define direction of the magnetic field. Dimensionality of
the Hilbert space of this system is $2s+1$. So an arbitrary state
of this system must be defined by $4s$ real parameters. It means
that we cannot provide evolution between two arbitrary states of a
spin-$s$ system with help of magnetic field. The quantum
brachistochrone evolution for a spin-1 system in a magnetic field
was considered in \cite{QBS1}.

Multilevel quantum systems with dimensionality higher than two
could be more efficient than qubit, because they provide a way for
more dense data recording. A three- and four-level systems are the
simplest multilevel systems after a two-level system. In quantum
information these systems are called {\it qutrit} and {\it
ququad}, respectively. In general, a $d$-level quantum system is
called {\it qudit}. The channel capacity for these systems is
greater than for a two-level system \cite{RCCLQOC}. The quantum
cryptography protocols created by qudits are more secure against
eavesdropping attacks than the cryptography protocols created by
qubits \cite{TQQCP, SQKDUDLS, QED1, QED2}. Therefore, qudits are
more efficient in many problems of quantum computation
\cite{PIUQP,QCBDLCS,AOQCDLS} and quantum cryptography \cite{QED1,
EQCQ}. Design of a qutrit quantum computer based on a trapped ion
in the presence of magnetic field gradient is presented in
\cite{TIQSMQC}. This work is the generalization of \cite{TIQSMQC1,
TIQSMQC2, TIQSMQC3}, where design of a qubit quantum computer on
trapped ion was considered. Another quantum system which is
suitable for quantum computations with qutrits is a polarized
biphoton \cite{QSEB, PSBQTL, OOCSMB, POB, SMBQPSR}, which is
formed by two correlated photons.

A geometric approach to study qudit system has been developed in
\cite{OCGQC, GAQCLB, QCAG, QGDM, GQCQ, BSSQWF, BVQUDIT, GQM,
RSMQS, CSRS}. In \cite{OCGQC, GAQCLB, QCAG, QGDM} it is shown that
in the case of qubit systems, finding an optimal quantum circuit
of a unitary operation is closely related  to the problem of
finding the minimal distance between two point on the Riemannian
metric. A similar problem for the case of $n$ qutrits was
considered in \cite{GQCQ}. The authors of this work showed that
the optimal quantum circuit is equivalent to the shortest path
between two points in a certain curved geometry of
$SU\left(3^n\right)$.

In \cite{BVQUDIT} the authors presented three different matrix
bases that can be used to decompose density matrices of a
$d$-dimensional quantum system. Namely, the generalized Gell-Mann
matrix basis, the polarization operator basis, and the Weyl
operator basis. These decompositions were identified with the
Bloch vector for qudit which is the generalization of the well
known qubit case. In \cite{GQM} it was shown that physical
characteristics of spin-$1/2$, spin-$1$, spin-$3/2$, and spin-$2$
systems can be represented by geometrical features that are
preferentially identified on the complex manifold.

The geometrical properties of some wellknown
coherent states manifolds, which are generated by an action of a
Lie group on a fixed states, was studied in details in
\cite{RSMQS, CSRS}. In these articles the Fubini-Study metric of
these manifolds was examined. The authors considered the atomic
coherent states, generated by the action of the $SU(2)$
displacement operator on the eigenstate of the $z$-component of
the angular momentum operator which corresponds to the lowest
eigenvalue. It was shown that the metric of the manifold of this
state is that of the sphere.

In this paper, we consider quantum brachistochrone evolution of
spin-$s$ system on the manifolds determined by a rotation of the
eigenstates of the operator of projection of spin-$s$ on some
direction. In Section \ref{sec2} it is shown that two such
manifolds exist for a spin-$1$ system. Each of them is defined by
two real parameters. Also, we show that they do not intersect each
other. The Fubini-Study metrics of these manifolds are obtained in
Section \ref{sec3}. It is shown that these are the metrics of the
spheres with radii dependent on the value of the spin and on the
value of the spin projection. The quantum brachistochrone problem
on each of the manifolds is considered in Section \ref{sec4}. We
generalize this problem for an arbitrary spin $s$ (Section
\ref{sec5}). In Section \ref{sec6} we give conclusions.

\section{The rotational manifolds of spin-$1/2$ and spin-$1$ systems \label{sec2}}

The rotation of the quantum state of spin-$s$ $\vert\psi_i\rangle$ through an angle $\chi$ about an axis in the direction of the unit vector ${\bf n}$ can be
realized as follows:
\begin{eqnarray}
\vert\psi_f\rangle=e^{-i\chi{\bf S}\cdot{\bf n}}\vert\psi_i\rangle,
\label{eq1}
\end{eqnarray}
where ${\bf S}$ is the spin-$s$ operator. In the spherical
coordinates the vector ${\bf n}$ can be represented as follows
$\bf{n}=\left(\sin\theta\cos\phi,\sin\theta\sin\phi,\cos\theta\right)$,
where $\theta$ and $\phi$ are the polar and azimuthal angles,
respectively. We set $\hbar=1$. For instance, let us consider the
rotation of the spin-$1/2$ system. In this case the spin-$1/2$
operator can be represented by the Pauli matrices
$\mbox{\boldmath{$ \sigma$}}$ as $\frac{1}{2}\mbox{\boldmath{$
\sigma$}}$. Rotations through an angle $\theta$ about the $y$-axis
and an angle $\phi$ about the $z$-axis allow us to achieve an
arbitrary quantum state of spin-$1/2$ having started from the
eigenvectors of $\sigma_z$
\begin{eqnarray}
&&\vert\psi^+\rangle=e^{-i\frac{\phi}{2}\sigma_z}e^{-i\frac{\theta}{2}\sigma_y}\vert\uparrow\rangle=\left( \begin{array}{ccccc} \cos\frac{\theta}{2}e^{-i\frac{\phi}{2}} \\ \sin\frac{\theta}{2}e^{i\frac{\phi}{2}} \end{array}\right),
\label{eq2}\\
&&\vert\psi^-\rangle=e^{-i\frac{\phi}{2}\sigma_z}e^{-i\frac{\theta}{2}\sigma_y}\vert\downarrow\rangle=\left( \begin{array}{ccccc} -\sin\frac{\theta}{2}e^{-i\frac{\phi}{2}} \\ \cos\frac{\theta}{2}e^{i\frac{\phi}{2}} \end{array}\right).
\label{eq3}
\end{eqnarray}
Here we use the fact that
\begin{eqnarray}
&&e^{-i\frac{\theta}{2}\sigma_{\alpha}}=\cos\frac{\theta}{2}-i\sin\frac{\theta}{2}\sigma_{\alpha},\nonumber
\end{eqnarray}
where $\alpha=x,y,z$. The states (\ref{eq2}) and (\ref{eq3}) are
the eigenstates of the operator $\mbox{\boldmath{$
\sigma$}}\cdot{\bf n}$ with $1$ and $-1$ eigenvalues,
respectively. Choosing parameters $\theta\in[0,\pi]$ and
$\phi\in[0,2\pi]$ in the equations either (\ref{eq2}) or
(\ref{eq3}) we can achieve an arbitrary state of the spin-$1/2$
system. In other words, these states cover the entire state space
of the spin-$1/2$ system.

Let us consider a similar problem for a spin-$1$ system. In the matrix representation components of the spin-$1$ operator read:
\begin{eqnarray}
S_x=\frac{1}{\sqrt{2}}\left( \begin{array}{ccccc}
0 & 1 & 0 \\
1 & 0 & 1 \\
0 & 1 & 0
\end{array}\right),\quad
S_y=\frac{i}{\sqrt{2}}\left( \begin{array}{ccccc}
0 & -1 & 0 \\
1 & 0 & -1 \\
0 & 1 & 0
\end{array}\right),\quad
S_z=\left( \begin{array}{ccccc}
1 & 0 & 0 \\
0 & 0 & 0 \\
0 & 0 & -1
\end{array}\right).
\label{eq4}
\end{eqnarray}
It is convenient to represent the operator which provides the
rotation of the quantum state of spin-$1$ around vector ${\bf n}$
in the form \cite{QBS1}
\begin{eqnarray}
e^{-i\chi{\bf S}\cdot{\bf n}} = 1-\left({\bf S}\cdot{\bf n}\right)^22\sin^2\frac{\chi}{2}-i{\bf S}\cdot{\bf n}\sin\chi .
\label{eq8}
\end{eqnarray}
The operator ${\bf S}\cdot{\bf n}$ has three eigenvalues $1$, $0$,
$-1$ with the corresponding eigenvectors $\vert\psi_1\rangle$,
$\vert\psi_0\rangle$, $\vert\psi_{-1}\rangle$. An arbitrary state
of a three-level system can be written as a linear combination of
these eigenvectors. It is enough to prove the equation (\ref{eq8})
only for these eigenvectors. It is easy to verify that for a
parameter $\lambda$, which takes only three values $1$, $0$ and
$-1$, we have
\begin{eqnarray}
e^{\lambda x} = (1-\lambda)(1+\lambda)+\frac{1}{2}\lambda(\lambda+1)e^x+\frac{1}{2}\lambda(\lambda-1)e^{-x}.
\label{eq9}
\end{eqnarray}
Then, using (\ref{eq9}) for the unitary operator of rotation, we
obtain the relation (\ref{eq8}). In general for the parameter
$\lambda$, which takes $n$ values, namely, $\lambda_1$,
$\lambda_2$, ..., $\lambda_n$, we have
\begin{eqnarray}
e^{\lambda x} = \sum_{m\neq k=1}^n\prod_{k=1}^n\frac{\lambda-\lambda_k}{\lambda_m-\lambda_k}e^{\lambda_mx}.
\label{eq10}
\end{eqnarray}

Now, using the equation (\ref{eq8}), we can represent the
operators which provide the rotations of the quantum state of the
spin-$1$ system around $x$-, $y$- and $z$-axis as follows
\begin{eqnarray}
e^{-i\theta S_{\alpha}} = 1-{S_{\alpha}}^22\sin^2\frac{\theta}{2}-iS_{\alpha}\sin\theta,\label{eq11}
\end{eqnarray}
where $\alpha=x,y,z$. The eigenstates of $S_z$ with the
eigenvalues $1$, $0$, $-1$ we denote as follows: $\vert 1\rangle$,
$\vert 0\rangle$, $\vert -1\rangle$.  These eigenvectors play the
role of the basis vectors. Let us consider the rotations of these
eigenstates through angles $\theta$ and $\phi$ about the $y$- and
$z$-axis, respectively. Then, using the equation (\ref{eq11}), we
obtain the following states
\begin{eqnarray}
&&\vert \psi_1\rangle=e^{-i\phi S_z}e^{-i\theta S_y}\vert 1\rangle = \left( \begin{array}{ccccc} \frac{1}{2}\left(1+\cos\theta\right)e^{-i\phi} \\ \frac{1}{\sqrt{2}}\sin\theta\\ \frac{1}{2}\left(1-\cos\theta\right)e^{i\phi} \end{array}\right),\label{eq5}\\
&&\vert \psi_0\rangle=e^{-i\phi S_z}e^{-i\theta S_y}\vert 0\rangle=\left( \begin{array}{ccccc} -\frac{1}{\sqrt{2}}\sin\theta e^{-i\phi} \\ \cos\theta\\ \frac{1}{\sqrt{2}}\sin\theta e^{i\phi} \end{array}\right),\label{eq6}\\
&&\vert \psi_{-1}\rangle=e^{-i\phi S_z}e^{-i\theta S_y}\vert -1\rangle=\left( \begin{array}{ccccc} \frac{1}{2}\left(1-\cos\theta\right)e^{-i\phi} \\ -\frac{1}{\sqrt{2}}\sin\theta\\ \frac{1}{2}\left(1+\cos\theta\right)e^{i\phi} \end{array}\right).\label{eq7}
\end{eqnarray}
It is important to note that these states are eigenstates of the
operator ${\bf S}\cdot{\bf n}$ with the corresponding eigenvalues
$1$, $0$ and $-1$, respectively. From the analysis of these
eigenstates it is clear that the states $\vert \psi_1\rangle$ and
$\vert \psi_{-1}\rangle$ belong to the same rotational manifold
and the state $\vert \psi_0\rangle$ belongs to another rotational
manifold. To cover the entire manifold defined by the states
$\vert \psi_1\rangle$ and $\vert \psi_{-1}\rangle$ it is enough
that the parameters $\theta$ and $\phi$ belong to the intervals
$\theta\in[0,\pi]$ and $\phi\in[0,2\pi]$. In the case of the
manifold defined by the state $\vert \psi_0\rangle$ we have that
it is twice covered by the intervals $\theta\in[0,\pi]$ and
$\phi\in[0,2\pi]$ because the following substitutions
$\theta\rightarrow\pi-\theta$ and $\phi\rightarrow\phi+\pi$
transform the state $\vert\psi_0\rangle$ into itself modulo a
global phase. At the same time these substitutions allow us to
transform the state $\vert\psi_1\rangle$ into the state
$\vert\psi_{-1}\rangle$. However, it does not exist any
substitution that transforms either the state $\vert
\psi_1\rangle$ or $\vert \psi_{-1}\rangle$ into the state $\vert
\psi_0\rangle$. It means that these manifolds do not intersect
each other.

In contrast to the case of the spin-$1/2$ system, where the
rotation manifold coincides with the two-dimensional quantum
space, none of the manifolds defined by the states
(\ref{eq5})-(\ref{eq7}) coincides with the quantum space of the
spin-$1$ system. The number of parameters which determine each of
these manifolds is not sufficient to specify the quantum space of
the spin-$1$ system which must be defined by four real parameters.
Moreover, a linear combination of the states which belong to one
of these manifolds does not belong to it.

\section{The Fubini-Study metrics of the rotational manifolds of spin-$1/2$ and spin-$1$ systems \label{sec3}}

The Fubini-Study metric is the infinitesimal distance $ds$ between two neighbouring pure quantum states $\vert\psi(\xi^{\alpha})\rangle$ and
$\vert\psi(\xi^{\alpha}+d\xi^{\alpha})\rangle$ \cite{FSM, FSM1, FSM2}. It is given by the following expression
\begin{eqnarray}
ds^2=g_{\alpha\beta}d\xi^{\alpha} d\xi^{\beta},
\label{eq16}
\end{eqnarray}
where $\xi^{\alpha}$ is a set of real parameters which define the state $\vert\psi(\xi^{\alpha})\rangle$. The components of the metric tensor
$g_{\alpha\beta}$ have the form:
\begin{eqnarray}
g_{\alpha\beta}=\gamma^2\Re\left(\langle\psi_{\alpha}\vert\psi_{\beta}\rangle-\langle\psi_{\alpha}\vert\psi\rangle\langle\psi\vert\psi_{\beta}\rangle\right),
\label{eq17}
\end{eqnarray}
where $\gamma$ is an arbitrary factor which is often chosen $1$, $\sqrt{2}$ or $2$ and
\begin{eqnarray}
\vert\psi_{\alpha}\rangle=\frac{\partial}{\partial\xi^{\alpha}}\vert\psi\rangle.
\label{eq18}
\end{eqnarray}

For instance, the Fubini-Study metric of the space of a spin-$1/2$
system, which is spanned by the states (\ref{eq2}) and
(\ref{eq3}), reads \cite{FSM,FSM2}
\begin{eqnarray}
ds^2=\frac{\gamma^2}{4}\left((d\theta)^2+\sin^2\theta(d\phi)^2\right).
\label{eq22}
\end{eqnarray}
Here, the angles $\theta$ and $\phi$ play the role of the
parameters $\xi^{\alpha}$. Note that (\ref{eq22}) is the metric of
the sphere of radius $\gamma/2$. In case of $\gamma=2$ we obtain
the metric of the Bloch sphere (the Bloch sphere is a sphere of
the unit radius which represents the state space of a two-level
system). The states $\vert\psi^+\rangle$ (\ref{eq2}) and
$\vert\psi^-\rangle$ (\ref{eq3}) correspond to the antipodal
points on this sphere.

Now let us calculate metrics of the rotational manifolds defined by the states (\ref{eq5})-(\ref{eq7}) obtained for $s=1$. These states are also determined by two real parameters
$\theta$ and $\phi$. As we mentioned earlier, the eigenstates (\ref{eq5}), (\ref{eq7}) belong to the same manifold and eigenstate (\ref{eq6}) belongs
to another manifold. Therefore, in order to obtain the Fubini-Study metric of these manifolds it is enough to consider the eigenstates (\ref{eq5}) and (\ref{eq6}).
Let us calculate the following derivatives from these eigenstates:
\begin{eqnarray}
&&\vert \psi_{1\ \theta}\rangle= \left(\begin{array}{ccccc} -\frac{1}{2}\sin\theta e^{-i\phi}\\ \frac{1}{\sqrt{2}}\cos\theta\\ \frac{1}{2}\sin\theta e^{i\phi} \end{array}\right),\quad
\vert \psi_{1\ \phi}\rangle= \left(\begin{array}{ccccc} -\frac{i}{2}\left(1+\cos\theta\right)e^{-i\phi}\\ 0\\ \frac{i}{2}\left(1-\cos\theta\right)e^{i\phi} \end{array}\right),\nonumber\\
&&\vert \psi_{0\ \theta}\rangle= \left(\begin{array}{ccccc} -\frac{1}{\sqrt{2}}\cos\theta e^{-i\phi}\\ - \sin\theta \\ \frac{1}{\sqrt{2}}\cos\theta e^{i\phi} \end{array}\right),\quad
\vert \psi_{0\ \phi}\rangle= \left(\begin{array}{ccccc} \frac{i}{\sqrt{2}}\sin\theta e^{-i\phi}\\ 0 \\ \frac{i}{\sqrt{2}}\sin\theta e^{i\phi}\end{array}\right).
\label{eq23}
\end{eqnarray}
Using these derivatives we obtain the following scalar products:
\begin{eqnarray}
&&\langle\psi_{1} \vert \psi_{1\ \theta}\rangle=0,\quad \langle\psi_{1\ \theta} \vert \psi_{1\ \theta}\rangle =\frac{1}{2},\nonumber\\
&&\langle\psi_{1} \vert \psi_{1\ \phi}\rangle=-i\cos\theta,\quad \langle\psi_{1\ \phi} \vert \psi_{1\ \phi}\rangle=\frac{1}{2}\left(1+\cos^2\theta\right),\nonumber\\
&&\langle\psi_{1\ \theta} \vert \psi_{1\ \phi}\rangle=\frac{i}{2}\sin\theta,\nonumber\\
&&\langle\psi_{0} \vert \psi_{0\ \theta}\rangle=0,\quad \langle\psi_{0\ \theta} \vert \psi_{0\ \theta}\rangle =1,\nonumber\\
&&\langle\psi_{0} \vert \psi_{0\ \phi}\rangle=0,\quad \langle\psi_{0\ \phi} \vert \psi_{0\ \phi}\rangle=\sin^2\theta,\nonumber\\
&&\langle\psi_{0\ \theta} \vert \psi_{0\ \phi}\rangle=0.
\label{eq24}
\end{eqnarray}
Substituting these products into the definition of the components of metric tensor (\ref{eq17}), the Fubini-Study metrics of the rotational manifolds defined by
the eigenstates (\ref{eq5})-(\ref{eq7}) take the form:
\begin{eqnarray}
&&ds^2_{1}=\frac{\gamma^2}{2}\left((d\theta)^2+\sin^2\theta(d\phi)^2\right),\label{eq25}\\
&&ds^2_{0}=\gamma^2\left((d\theta)^2+\sin^2\theta(d\phi)^2\right),\label{eq26}
\end{eqnarray}
where subscript in $ds$ indicates the eigenvalue that in turn
indicates the manifold. It is easy to see that the expression
(\ref{eq25}) describes metric of the sphere of radius
$\gamma/\sqrt{2}$. The orthogonal states correspond to antipodal
points on this sphere. In the case of manifold which is defined by
(\ref{eq26}) we obtain another result. As we mentioned earlier,
the substitutions $\theta\rightarrow\pi-\theta$ and
$\phi\rightarrow\phi+\pi$ transform the state $\vert\psi_0\rangle$
(\ref{eq6}) into itself modulo a global phase. The manifold
defined by this state is called elliptic geometry. It is important
to note that orthogonal states on the manifold (\ref{eq26}) are
separated by an angle $\pi/2$. Really, the scalar product of two
states $\vert\psi_0\rangle\equiv\vert\psi_0 (\theta,\phi)\rangle$
and $\vert\psi_0'\rangle\equiv\vert\psi (\theta',\phi')\rangle$,
which belong to manifold (\ref{eq26}), reads
\begin{eqnarray}
\langle\psi_0\vert\psi_0'\rangle=\sin\theta\sin\theta'\cos(\phi-\phi')+\cos\theta\cos\theta'.
\label{eq28}
\end{eqnarray}
On the other hand, this is the scalar product of two unit vectors
${\bf n}$ defined by the spherical angles $\theta$, $\phi$ and
${\bf n'}$ defined by the spherical angles $\theta'$, $\phi'$,
respectively. This product becomes zero when the angle between
these vectors is $\pi/2$. This confirms our conclusion.

\section{The quantum brachistochrone problem for spin-$1$ system in a magnetic field \label{sec4}}

In this section we consider quantum brachistochrone evolution on
the rotational manifolds defined by the metrics (\ref{eq25}),
(\ref{eq26}), obtained for spin-$1$ system. Hamiltonian providing
such evolution is the Hamiltonian of the spin-$1$ system in an
external magnetic field directed along the unit vector ${\bf n'}$
\begin{eqnarray}
H=\omega{\bf S}\cdot{\bf n'},
\label{eq29}
\end{eqnarray}
where $\omega$ is proportional to the strength of the magnetic
field and is measured in frequency units, ${\bf n'}$ is defined by
two spherical angles $\theta'$ and $\phi'$. As we mentioned above
we set $\hbar=1$. The eigenvalues of this Hamiltonian are
$\omega$, $0$ and $-\omega$ with the corresponding eigenstates
(\ref{eq5})-(\ref{eq7}), where angles $\theta$ and $\phi$ are
denoted as $\theta'$ and $\phi'$, respectively. Hamiltonian
(\ref{eq29}) contains only two free parameters, namely, two angles
$\theta'$ and $\phi'$. The general state for a spin-$1$ system is
defined by four real parameters. Therefore, we cannot reach an
arbitrary state using the operator of evolution with Hamiltonian
(\ref{eq29}).

The quantum brachistochrone problem for spin-$1$ system in a
magnetic field is studied in the paper \cite{QBS1}. The authors
considered the following question: what is the optimal direction
of the magnetic field ${\bf n'}$ at the fixed value $\omega$, such
that the evolution from a given initial state $\vert\psi_i\rangle$
to a given final one $\vert\psi_f\rangle$ is achieved in the
shortest time? In that paper, studying directly the evolution of
quantum state with the Hamiltonian (\ref{eq29}), conditions for
optimal evolution were obtained. We solve this problem using
geometric properties of manifolds defined by (\ref{eq25}) and
(\ref{eq26}). Let us consider it in detail.

Using equation (\ref{eq9}), the unitary operator of evolution with Hamiltonian (\ref{eq29}) takes the form
\begin{eqnarray}
e^{-iHt}=1-\left({\bf S}\cdot{\bf n'}\right)^22\sin^2\frac{\omega t}{2}-i{\bf S}\cdot{\bf n'}\sin\omega t.
\label{eq30}
\end{eqnarray}
Now, using this operator we can consider the quantum evolution of the system described by Hamiltonian (\ref{eq29}).
Let us take the initial states as the eigenstates of $S_z$ $\vert 1\rangle$, $\vert 0\rangle$ and $\vert -1\rangle$ \cite{QBS1}.
Then using (\ref{eq30}), we finally find
\begin{eqnarray}
\vert\psi_1(t)\rangle=e^{-iHt}\vert 1\rangle =\left(\begin{array}{ccccc} 1-\left(1+\cos^2\theta'\right)\sin^2\frac{\omega t}{2}-i\cos\theta'\sin\omega t \\
-\left(\sqrt{2}\cos\theta'\sin\theta'\sin^2\frac{\omega t}{2}+\frac{i}{\sqrt{2}}\sin\theta'\sin\omega t\right)e^{i\phi'}\\
-\sin^2\theta'\sin^2\frac{\omega t}{2}e^{i2\phi'} \end{array}\right),\label{eq31}\ \ \ \ \ \ \
\end{eqnarray}
\begin{eqnarray}
\vert\psi_0(t)\rangle=e^{-iHt}\vert 0\rangle =\left(\begin{array}{ccccc} -\frac{1}{\sqrt{2}}\left(2\cos\theta'\sin\theta'\sin^2\frac{\omega t}{2}+i\sin\theta'\sin\omega t\right)e^{-i\phi'}\\
1-2\sin^2\theta'\sin^2\frac{\omega t}{2}\\
\frac{1}{\sqrt{2}}\left(2\cos\theta'\sin\theta'\sin^2\frac{\omega t}{2}-i\sin\theta'\sin\omega t\right)e^{i\phi'} \end{array}\right),\label{eq32}\ \ \ \ \ \ \ \ \\
\vert\psi_{-1}(t)\rangle=e^{-iHt}\vert -1\rangle =\left(\begin{array}{ccccc} -\sin^2\theta'\sin^2\frac{\omega t}{2}e^{-i2\phi'}\\
\left(\sqrt{2}\cos\theta'\sin\theta'\sin^2\frac{\omega t}{2}-\frac{i}{\sqrt{2}}\sin\theta'\sin\omega t\right)e^{-i\phi'}\\
1-\left(1+\cos^2\theta'\right)\sin^2\frac{\omega t}{2}+i\cos\theta'\sin\omega t \end{array}\right).\label{eq33}\ \ \
\end{eqnarray}
It is easy to show that the states (\ref{eq31})-(\ref{eq33}) are
equal to the eigenstates (\ref{eq5})-(\ref{eq7}) modulo a global
phase:
\begin{eqnarray}
&&\vert \psi_1(t)\rangle= e^{i\beta}\vert\psi_1\rangle,\label{eq35}\\
&&\vert \psi_0(t)\rangle= \vert\psi_0\rangle,\label{eq36}\\
&&\vert \psi_{-1}(t)\rangle= e^{-i\beta}\vert\psi_{-1}\rangle,\label{eq37}
\end{eqnarray}
where
\begin{eqnarray}
\beta=2\phi'-\phi+(2k+1)\pi,\nonumber
\end{eqnarray}
$k$ is an arbitrary integer. Here we introduce the following notation:
\begin{eqnarray}
&&\phi=\phi'-\arctan\frac{\cos\frac{\omega t}{2}}{\cos\theta'\sin\frac{\omega t}{2}},\label{eq341}\nonumber\\
&&\sin\frac{\theta}{2}=\sin\theta'\sin\frac{\omega t}{2}.
\label{eq34}
\end{eqnarray}
As we can see, if the initial state belongs to one of these
manifolds then the quantum evolution of the system takes place on
the same manifold. In other words, as we mentioned previously,
Hamiltonian (\ref{eq29}) realizes quantum evolution on two
manifolds separately and does not mix them. For instance, we
cannot achieve the state $\vert 0\rangle$ starting from the state
$\vert 1 \rangle$.

\begin{figure*}[h]
    \centering
        \includegraphics[width=0.5\textwidth]{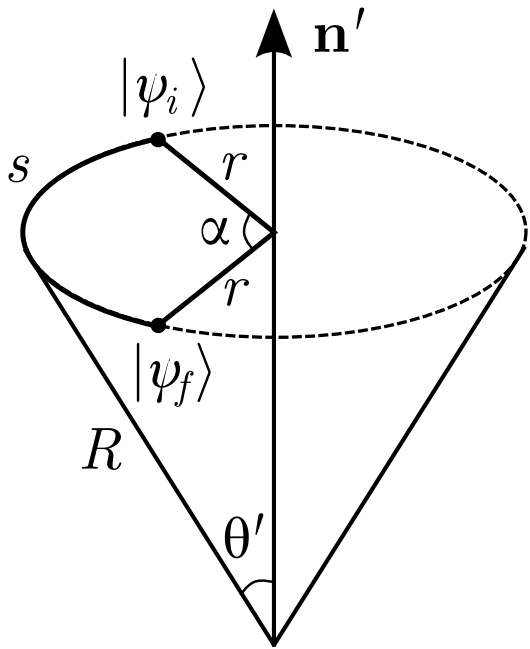}
    \caption{\textit{Identification of the path length of evolution $s$ for a spin system in the magnetic field directed along the unit vector {\bf n}'.}}
    \label{brach}
\end{figure*}
The period of time of evolution from the initial
state $\vert\psi_i\rangle$ to the final one $\vert\psi_f\rangle$
is given by the ratio
\begin{eqnarray}
t=\frac{s}{v},
\label{eq341}
\end{eqnarray}
where $s$ is the path length of evolution between these states and
$v$ is the speed of evolution. The shortest path joining the two
states on the sphere is the length of the great circle arc (the
length of the geodesic).

Using results obtained above, let us consider quantum
brachistochrone problem on the manifolds defined by (\ref{eq25}),
(\ref{eq26}), separately. First, we examine the optimal evolution
on the manifold described by equation (\ref{eq25}). We take the
initial and the final states which belong to the manifold defined
by the metric (\ref{eq25}) as follows $\vert\psi_i\rangle=\vert
1\rangle$,
\begin{eqnarray}
\vert \psi_f\rangle= \left(\begin{array}{ccccc} \frac{1}{2}\left(1+\cos\theta_f\right)e^{-i\phi_f} \\ \frac{1}{\sqrt{2}}\sin\theta_f \\ \frac{1}{2}\left(1-\cos\theta_f\right)e^{i\phi_f} \end{array}\right).
\label{eq38}
\end{eqnarray}
The final state is achieved when the angle between
the magnetic field and this state is the same as the angle between
the magnetic field and the initial state (see Fig. \ref{brach}).
Then the quantum evolution between two states $\vert\psi_i\rangle$
and $\vert\psi_f\rangle$ takes place along the arc of a circle $s$
around the unit vector ${\bf n'}$. From the analysis of the Fig.
\ref{brach} it is clear that
\begin{eqnarray}
s=\alpha r,
\label{eq381}
\end{eqnarray}
where $r=R\sin\theta'$. The initial and the final states are
separated by an angle $\theta_f$, therefore the angle $\alpha$
takes the form
$\alpha=2\arcsin\frac{\sin\frac{\theta_f}{2}}{\sin\theta'}$.
Substituting this expression into (\ref{eq381}), we obtain the
path length of evolution between the initial and the final states
on the manifold defined by (\ref{eq25}) as follows
\begin{eqnarray}
s=2R\sin\theta'\arcsin\frac{\sin\frac{\theta_f}{2}}{\sin\theta'},
\label{eq382}
\end{eqnarray}
where $R=\frac{\gamma}{\sqrt{2}}$ is the radius of the manifold being a sphere.

Now it is necessary to calculate the speed of evolution between $\vert\psi_i\rangle$ and $\vert\psi_f\rangle$ states.
The speed $v$ of quantum evolution is given by the Anandan-Aharonov relation \cite{AA} as
\begin{eqnarray}
v=\gamma\sqrt{\langle\psi (t)\vert\left(\Delta H\right)^2\vert\psi (t)\rangle}.
\label{eq40}
\end{eqnarray}
In order to calculate the speed of evolution on the manifold
defined by the equation (\ref{eq25}) let us substitute the state
(\ref{eq31}) and the Hamiltonian (\ref{eq29}) into the equation
(\ref{eq40})
\begin{eqnarray}
v=\gamma\sqrt{\langle\psi_1(t)\vert\left(\Delta H\right)^2\vert\psi_1(t)\rangle}=\gamma\sqrt{\langle 1\vert\left(\Delta H\right)^2\vert 1\rangle}=\omega R\sin\theta'.
\label{eq41}
\end{eqnarray}
Then, using the equation (\ref{eq341}) with
equations (\ref{eq382}) and (\ref{eq41}), we find the period of
time of evolution between the initial state
$\vert\psi_i\rangle=\vert 1\rangle$ and the final one (\ref{eq38})
\begin{eqnarray}
t=\frac{2}{\omega}\arcsin\frac{\sin\frac{\theta_f}{2}}{\sin\theta'}.
\label{eqtime}
\end{eqnarray}
The minimal period of time is achieved for
$\theta'=\frac{\pi}{2}$. We have
\begin{eqnarray}
t_{min}=\frac{\theta_f}{\omega}.
\label{eq43}
\end{eqnarray}
This condition corresponds to the minimal length
of path $s_{min}=\gamma\theta_f/\sqrt{2}$ and the maximal speed of
evolution $v_{max}=\gamma\omega/\sqrt{2}$. For example, in the
case of $\theta_f=\pi$, the minimal path and the minimal time of
evolution between two orthogonal states read
$s_{min}=\gamma\pi/\sqrt{2}$ and $t_{min}=\pi/\omega$. So, the
optimal evolution is achieved for perpendicular orientation of the
magnetic field with respect to the initial and the final states.
It means that the unit vector which defines
direction of the magnetic field takes the form ${\bf
n}'_{opt}=\left(-\sin\phi_f,\cos\phi_f,0\right)$. The Hamiltonian
which provides the optimal evolution takes the following form
\begin{eqnarray}
H_{opt}=\omega{\bf S}\cdot{\bf n'_{opt}}.
\label{eqhamopt}
\end{eqnarray}

The same situation we have in the case of the manifold defined by
the equation (\ref{eq26}). Here we consider evolution between the
initial state $\vert\psi_i\rangle=\vert 0\rangle$ and the final
one
\begin{eqnarray}
\vert \psi_f\rangle= \left(\begin{array}{ccccc} -\frac{1}{\sqrt{2}}\sin\theta_f e^{-i\phi_f} \\ \cos\theta_f  \\ \frac{1}{\sqrt{2}}\sin\theta_f e^{i\phi_f}\end{array}\right).\nonumber
\end{eqnarray}
Then, having performed the same steps as in the previous case, we
obtain that the length of the path which the system passes between
these states and the speed of evolution of the system are also
defined by the equations (\ref{eq382}) and (\ref{eq41}),
respectively. But here the manifold has the following radius
$R=\gamma$. As we can see, similarly to the previous case, the
period of evolution is defined by the equation (\ref{eqtime}). In
this case the optimal evolution also corresponds to the
perpendicular orientation of the magnetic field to the initial and
the final states. Hence the optimal period of time is defined by
the equation (\ref{eq43}) and the Hamailtonian which provides such
evolution is defined by the equation (\ref{eqhamopt}). Here the
minimal time of evolution between these two orthogonal states is
$\pi/2\omega$ because they are separated by an angle $\pi/2$.

\section{Generalization for an arbitrary spin \label{sec5}}

The problem which we considered in the previous sections can be
generalized for an arbitrary spin. Namely, what are the geometries
of the rotational manifolds which determine the position of the
states achieved by the rotation of the eigenstates of the operator
of projection of spin-$s$ on the direction ${\bf n}$? The
eigenstate of the operator ${\bf S}\cdot{\bf n}$ with an
eigenvalue $m$ can be represented as follows
\begin{eqnarray}
\vert \psi_m\rangle = e^{-i\phi S_z}e^{-i\theta S_y}\vert m\rangle,
\label{eq54}
\end{eqnarray}
where ${\bf S}$ is the operator of spin-$s$, ${\bf n}$ is defined
by the spherical angles $\theta$ and $\phi$, $\vert m\rangle$ is
the eigenstate of $S_z$ with the eigenvalue $m$. Here the
eigenstates of $S_z$ play the role of the basis vectors. As we can
see, the eigenstate (\ref{eq54}) is defined by two real parameters
$\theta$ and $\phi$. It is rather difficult to represent the
eigenstates of the operator ${\bf S}\cdot{\bf n}$ for spin-$3/2$
in the ordinary form and to perform calculations for these states.
Therefore, to simplify further calculations we will use the
eigenstates of the operator ${\bf S}\cdot{\bf n}$ written in the
form (\ref{eq54}).

To obtain metric of the manifold defined by the state (\ref{eq54})
let us calculate the following derivatives
\begin{eqnarray}
&&\vert \psi_{m\ \theta}\rangle= e^{-i\phi S_z}\left(-iS_y\right)e^{-i\theta S_y}\vert m\rangle,\nonumber\\
&&\vert \psi_{m\ \phi}\rangle= \left(-iS_z\right)e^{-i\phi S_z}e^{-i\theta S_y}\vert m\rangle.
\label{eq55}
\end{eqnarray}
Then we can write the following scalar products
\begin{eqnarray}
&&\langle\psi_m \vert \psi_{m\ \theta}\rangle=-i\langle m \vert  S_y\vert m\rangle,\label{eq56}\\
&&\langle\psi_{m\ \theta} \vert \psi_{m\ \theta}\rangle=\langle m \vert  {S_y}^2\vert m\rangle,\label{eq57}\\
&&\langle\psi_m \vert \psi_{m\ \phi}\rangle =-i\langle m \vert e^{i\theta S_y} S_ze^{-i\theta S_y}\vert m\rangle,\label{eq58}\\
&&\langle\psi_{m\ \phi} \vert \psi_{m\ \phi}\rangle=\langle m \vert e^{i\theta S_y} {S_z}^2e^{-i\theta S_y}\vert m\rangle,\label{eq59}\\
&&\langle\psi_{m\ \theta} \vert \psi_{m\ \phi}\rangle=\langle m \vert e^{i\theta S_y} S_yS_ze^{-i\theta S_y}\vert m\rangle.
\label{eq60}
\end{eqnarray}
Having calculated these scalar products we obtain
\begin{eqnarray}
&&\langle\psi_m \vert \psi_{m\ \theta}\rangle=0,\quad \langle\psi_{m\ \theta} \vert \psi_{m\ \theta}\rangle=\frac{1}{2}\left(s+s^2-m^2\right)\nonumber\\
&&\langle\psi_m \vert \psi_{m\ \phi}\rangle =-im\cos\theta,\nonumber\\
&&\langle\psi_{m\ \phi} \vert \psi_{m\ \phi}\rangle=\frac{1}{2}\left(s+s^2-m^2\right)\sin^2\theta+m^2\cos^2\theta,\nonumber\\
&&\langle\psi_{m\ \theta} \vert \psi_{m\ \phi}\rangle=i\frac{m}{2}\sin\theta.
\label{eq601}
\end{eqnarray}
Here we use Baker-Campbell-Hausdorff formula for following operators
\begin{eqnarray}
&&e^{i\theta S_y} S_ze^{-i\theta S_y}=S_z\cos\theta-S_x\sin\theta,\nonumber\\
&&e^{i\theta S_y} {S_z}^2e^{-i\theta S_y}=e^{i\theta S_y} S_z e^{-i\theta S_y} e^{i\theta S_y}S_ze^{-i\theta S_y}=\left(e^{i\theta S_y} S_ze^{-i\theta S_y}\right)^2\nonumber\\
&&={S_x}^2\sin^2\theta+{S_z}^2\cos^2\theta-\left(S_xS_z+S_zS_x\right)\cos\theta\sin\theta.\nonumber
\end{eqnarray}
Substituting these scalar products into the equation (\ref{eq17}),
we obtain the components of the metric tensor as follows
\begin{eqnarray}
g_{\theta\theta}=\frac{\gamma^2}{2}\left(s+s^2-m^2\right),\quad
g_{\phi\phi}=\frac{\gamma^2}{2}\left(s+s^2-m^2\right)\sin^2\theta,\quad
g_{\theta\phi}=0. \label{eq61}
\end{eqnarray}
Thus, the Fubini-Study metric of the manifold defined by state (\ref{eq54}) is
\begin{eqnarray}
ds^2_m=\frac{\gamma^2}{2}\left(s+s^2-m^2\right)\left((d\theta)^2+\sin^2\theta(d\phi)^2\right).
\label{eq62}
\end{eqnarray}
As we see this is the metric of the sphere of radius
\begin{eqnarray}
R=\frac{\gamma}{\sqrt{2}}\sqrt{s+s^2-m^2}.
\label{eq622}
\end{eqnarray}
The Fubini-Study metric of the manifold with $m=-s$ was considered in \cite{RSMQS, CSRS}. We obtain the
metric for rotational manifolds with arbitrary $m$.

From the analysis of (\ref{eq62}) it is clear that there exist $s+1$ manifolds for an integer spin and $s+1/2$ manifolds for a half-integer spin.
For instance, in the case of spin-$3/2$ system we have two rotational manifolds with radii $\gamma\sqrt{3}/2$ and $\gamma\sqrt{7}/2$
which correspond $m=\pm 3/2$ and $m=\pm 1/2$, respectively. The same result we obtain directly, using the ordinary form of the eigenstates of the operator
${\bf S}\cdot{\bf n}$ for spin-$3/2$.

Let us consider the evolution of spin-$s$ system which takes place
on the manifold defined by the metric (\ref{eq62}). Hamiltonian
which allows us to provide such evolution has the form
(\ref{eq29}) with the spin-$s$ operator. We take the initial state
as follows $\vert\psi_i\rangle=\vert m\rangle$ and the final one
as follows $\vert\psi_f\rangle=e^{-i\phi_f S_z}e^{-i\theta_f
S_y}\vert m\rangle $. Making the same steps as in the case of
spin-$1$ system, we obtain that the length of path between the
initial and final states is defined by the equation (\ref{eq382})
with radius (\ref{eq622}). Using the Anandan-Aharonov relation
(\ref{eq40}) with the Hamiltonian (\ref{eq29}) for spin-$s$
system, we obtain that in the general case the speed of evolution
depends on the radius of manifold (\ref{eq622}) and on the
direction of the magnetic field as follows
\begin{eqnarray}
&&v=\gamma\sqrt{\langle\psi_m(t)\vert\left(\Delta H\right)^2\vert\psi_m(t)\rangle}=\gamma\sqrt{\langle m\vert\left(\Delta H\right)^2\vert m\rangle}\nonumber\\
&&=\omega R\sin\theta'.
\label{eq651}
\end{eqnarray}
So, similarly to the case of spin-$1$ system the optimal evolution happens when the magnetic field is directed perpendicular
to the initial and the final states. Then the shortest path between two states which are separated by the angle $\theta_f$ is
\begin{eqnarray}
s_{min}=\theta_fR=\theta_f\frac{\gamma}{\sqrt{2}}\sqrt{s+s^2-m^2}
\label{eq63}
\end{eqnarray}
and the maximal speed is
\begin{eqnarray}
v_{max}=\omega R=\omega\frac{\gamma}{\sqrt{2}}\sqrt{s+s^2-m^2}.
\label{eq64}
\end{eqnarray}
Then, using (\ref{eq341}) with (\ref{eq63}) and (\ref{eq64}), we obtain that the minimal time of evolution between two states separated by angle $\theta_f$
is determined by equation (\ref{eq43}). The optimal Hamiltonian which provides such evolution is defined by (\ref{eqhamopt}) with the spin-$s$ operator.

\section{Conclusion \label{sec6}}

Rotations of the eigenstate of the operator $S_z$ with eigenvalue
$m$ through an angle $\theta$ about the $y$-axis and an angle
$\phi$ about the $z$-axis allow us to achieve the eigenstate of
the operator of projection of spin-$s$ on the direction ${\bf
n}(\theta,\phi)$ with the same eigenvalue. This eigenstate belongs
to some manifold called rotation manifold defined by two real
parameters $\theta$ and $\phi$. For a spin-$1/2$ system there
exists one rotational manifold which coincides with a
two-dimensional quantum space. In the general case, there exist
$s+1$ manifolds for an integer spin and $s+1/2$ manifolds for a
half-integer one. The rotational manifolds for an arbitrary
spin-$s$ system (excluding the case with spin-$1/2$ system) do not
coincide with the quantum space of this system. The number of
parameters defining each those manifolds is not enough to specify
the quantum space of a spin-$s$ system which is given by $4s$ real
parameters. Moreover, rotational manifolds for a spin greater than
$1/2$ do not have properties of linear spaces. Linear combination
of the states which belong to one of such manifolds does not
belong to it.

For the spin-$1$ system it was shown that there are two rotational
manifolds which correspond to the eigenvalues $m=\!\pm 1$ and
$m\!=0$ of the operator ${\bf S}\cdot{\bf n}$. The Fubini-Study
metric of the manifold for $m=\pm 1$ is that of the sphere of
radius $\gamma/\sqrt{2}$. The orthogonal states correspond to
antipodal points on this sphere. The Fubini-Study metric of the
another manifold (with $m=0$) is that of the sphere of radius
$\gamma$. We showed that this manifold has properties of elliptic
geometry. Here orthogonal states are separated by the angle
$\pi/2$. These results were generalized for the arbitrary spin
$s$. In the general case the Fubini-Study metric of the rotational
manifold which corresponds to the eigenvalue $m$ is that of the
sphere with the radius dependent on the value of the spin $s$ and
on the value of the spin projection $m$ and is defined by the
equation (\ref{eq622}). In \cite{RSMQS, CSRS} the
Fubini-Study metric of the manifold was considered only for
particular case when $m=-s$. We want to emphasize that we obtained
metric for rotational manifolds for arbitrary $m$.

Finally, we considered quantum evolution for the spin-$1$ system
on the rotational manifolds. We solved the quantum brachistochrone
problem for the spin-$1$ system in the magnetic field using
geometric properties of the rotational manifolds. We conclude that
the optimal evolution happens when the magnetic field is
perpendicular to the initial and the final states.
The Hamiltonian which provides such evolution is
defined by the equation (\ref{eqhamopt}). Then the minimal path
length between these states which are separated by an angle
$\theta_f$ is a geodesic line on the rotational manifolds. The
minimal period of time of evolution on these manifolds is defined
by the equation (\ref{eq43}). We obtained similar results in the
general case of a spin-$s$ system. Namely, we obtained that the
minimal period of time of evolution of a spin-$s$ system in the
magnetic field between the initial and the final states separated
by an angle $\theta_f$ is defined by the equation (\ref{eq43}).
The Hamiltonian which provides optimal evolution
of spin-$s$ system is defined by the equation (\ref{eqhamopt})
with spin-$s$ operator. We note that the Hamiltonian of the
spin-s system in the magnetic field (excluding the case of
spin-$1/2$ system) does not contain enough number of parameters to
provide the evolution between two arbitrary quantum states.
Therefore, we could not consider the quantum brachistochrone
problem for this system on the whole Hilbert space.

\section{Acknowledgment}

The authors would like to thank Dr. Andrij Rovenchak, Dr. Mykola
Stetsko, Yuri Krynytskyi,  and Khrystyna Gnatenko for useful
comments.

\end{document}